\documentclass[10pt,A4paper,conference]{./IEEEtran}
\usepackage{epsfig,graphics}
\usepackage{amsmath,pifont,amssymb,amsfonts}

\newenvironment{matrice}
{\left[\begin{array}}{\end{array}\right]}

\newcommand{\field}[1]{\mathbb{#1}}

\begin{document}

 \title{A Closed-Form Solution for the Finite Length Constant Modulus Receiver }

\author{\authorblockN{Christophe Laot and Nicolas Le Josse}
\authorblockA{Dept. of Signal and Communications, TAMCIC (CNRS 2658)\\
Technopôle de Brest Iroise, CS 83818 29238 BREST Cedex\\
Email: christophe.laot@enst-bretagne.fr, nicolas.lejosse@enst-bretagne.fr}}

\maketitle

\begin{abstract}

In this paper, a closed-form solution minimizing the Godard or Constant Modulus (CM) cost function under the practical conditions of finite SNR and finite equalizer length  is derived. While previous work has been reported by Zeng \textit{et al.}, \textit{IEEE Trans. Information Theory}. 1998, to establish the link between the constant modulus and Wiener receivers, we show that under the Gaussian approximation of intersymbol interference at the output of the equalizer, the CM finite-length receiver is equivalent to the nonblind MMSE equalizer up to a complex gain factor. Some simulation results are provided to support the Gaussian approximation assumption.

\end{abstract}


\section{Introduction}
The transmission of information over frequency selective digital communications channels is subject to intersymbol interference (ISI). Equalization has proven to be an effective means for removing ISI. The commonly used approach employs a known sequence of training symbols and the equalizer coefficients are then adapted by using some adaptive algorithm so that the output of the equalizer closely matches the training sequences by minimizing the mean-squares-error (MMSE) criterion. However, in many high data rate bandlimited digital communications systems, the transmission of a training sequence results in a significant reduction in the effective communication link data rate. Therefore, instead of using a training sequence, only some statistical or structural properties of the transmitted and received signal can be exploited in a blind process to adapt the equalizer i.e. blind equalization. One of the simplest and the most effective blind equalization schemes is the constant modulus algorithm (CMA) \cite{Godard-1,Treichler-4}, which uses constant modularity as the desired property of the output of the receiver. A great deal of research about convergence behavior has been reported but a lack of comprehension about the global convergence of these algorithms has limited their utilization. Global convergence has been proven under ideal conditions  \cite{Godard-1,Treichler-4,Shalvi-2,Benveniste-3} (e.g., noise-free and doubly infinite equalizer) providing the inverse transfer function of the channel. However, this assumption can be excessive and the channel noise must be considered in the design of the equalizer. Recently, Zeng \textit{et al.} \cite{Zeng-6} have proposed interesting results on the relashionships between blind and Wiener receivers in the noisy case. They  observe that if a Wiener equalizer reaches an acceptable MSE performance, there exists a CM equalizer in its immediate neighborhood which is approximately a scaled version of the Wiener receiver. Nevertheless, never has a closed-form solution for the CM equalizer been proposed in the noisy case. In this paper, a closed-form solution for the finite-length CM equalizer for a non-minimum phase channel in the presence of additive white Gaussian noise is given. The CM equalizer is shown to be equivalent to the MMSE equalizer up to a complex gain factor. In order to illustrate the theoretical result, a simulation for 4PSK signals is performed, where a comparison of MMSE and CM equalizers, using gradient stochastic algorithms, is implemented.

\section{ A modified expression of the CM cost function }
We consider a discrete-time baseband transmission model.  Complex symbols $a_n$ are independent, identically distributed (\textit{i.i.d.}) random variables with zero-mean, variance $\sigma_a^2$ and sub-Gaussian distribution ($K(a)<0$). Symbols are sent through a discrete-time equivalent channel with coefficients $h_l$. The output of the channel is corrupted by ISI and can be expressed by
	\begin{equation}
	y_n = \sum_{l=0}^{L} h_l a_{n-l} + b_n
	\label{equ:model_siso}
	\end{equation}
where ${b_n}$ is a zero-mean, white Gaussian complex and circularly-symmetric random process with variance $\sigma_b^2$ and $L$ is the channel memory. Considering a block of $M$ output symbols and assuming that the channel is time-invariant over this block, (\ref{equ:model_siso}) can be written more compactly as follows
	\begin{equation}
	Y_n= HA_{n} + B_n
	\label{equ:model_matrix}
	\end{equation}
 where we define the column vector $\begin{array}{ccl}
A_{n}=
\begin{matrice}{cccc}
a_{n}&a_{n-1}&\hdots &a_{n-(M+L-2)}
\end{matrice}
^{T}
\end{array}$ $\in {\field{C}}^{M+L-1}$  of the input symbol sequence and the additive noise vector $\begin{array}{ccl}
B_{n}=
\begin{matrice}{cccc}
b_{n}&b_{n-1}&\hdots &b_{n-(M-1)}
\end{matrice}
^{T}
\end{array}$  $\in {\field{C}}^{M}$.
$H$ is the ${M\times (M+L-1) }$ channel Toeplitz matrix. Denote superscript $^{T}$ as the transpose operator.
If $\begin{array}{ccl}
C=
\begin{matrice}{cccc}
c_{0}&c_{1}&\hdots&c_{M-1}
\end{matrice}
^{T}
\end{array}$
 $\in {\field{C}}^{M}$  is the tap-weight vector of the equalizer of length $M$, the output of the equalizer ${z_n}$ can be represented by
      \begin{equation}
	z_n =C^{H}Y_{n}
	\label{equ:model_equal}
	\end{equation}
where superscript $^{H}$ is the Hermitian transpose operator. The CM cost function \cite{Godard-1} with index $p=2$ is defined by 
	\begin{equation}
	D^{(2)}=E\{(|z_n|^2-R_2)^2\}
	\label{equ:cost_function_CMeq3}
	\end{equation}
where $R_2$ is a real positive dispersion factor.
A necessary condition for the minimization of (\ref{equ:cost_function_CMeq3}) is given by
	\begin{equation}
	\nabla_{c} D^{(2)}=E\{Y_{n}Y_{n}^{H}C(|z_n|^2-R_2)\}=0 
	\label{equ:minim_CM}
	\end{equation}
where we define the gradient of $D^{(2)}$ at $C$ as the row vector $\nabla_{\mathbf{c}} D^{(2)}$.
Then, one can write the following relation
	\begin{equation}
	C^{H}\nabla_{c} D^{(2)}=E\{|z_n|^2(|z_n|^2-R_2)\}=0
	\label{eq:minim_CM2}
	\end{equation}
Finally, the condition necessary to reach a minimum of the CM cost function involves the equality between the second and fourth order moments of $z_n$ up to a positive gain $R_2$
	\begin{equation}
	E\{|z_n|^4\}=R_2 E\{|z_n|^2\}
	\label{eq:eq7}
	\end{equation}
Considering now the development of (\ref{equ:cost_function_CMeq3}), we have
	\begin{equation}
	D^{(2)}=R_2(R_2-E\{|z_n|^2\}) + E\{|z_n|^4\} - R_2 E\{|z_n|^2\} 
	\label{eq:eq8}
	\end{equation}
Then, a cost function equivalent to $D^{(2)}$ subject to the necessary condition (\ref{eq:eq7}) can be defined by
	\begin{equation}
	\bar{D}^{(2)}=R_2(R_2-E\{|z_n|^2\})
	\label{eq:eq9}
	\end{equation}
Thus, the minimization of $\bar{D}^{(2)}$ under the constraint (\ref{eq:eq7}) is equivalent to the minimization of the CM cost function $D^{(2)}$. 

\section{ Finite-length equalizer minimizing the CM cost function }
\begin{figure}[htb]
\centering
\includegraphics[height=3cm,width=8cm]{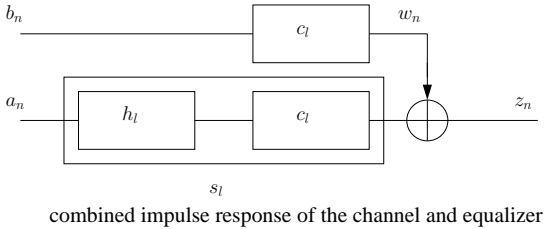}
\caption{Channel- equalizer block diagram. The cascade channel-equalizer is known as the global response.}\label{fig:block_diagram}
\end{figure}

\begin{figure}[htb]
\centering
\includegraphics[height=2.5cm,width=9cm]{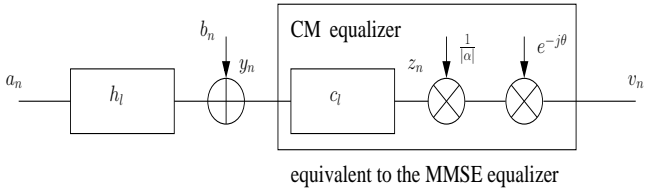}
\caption{CM equalizer equivalent to the MMSE equalizer.}\label{fig:comm_model_MMSE}
\end{figure}

From (\ref{equ:model_matrix}) and (\ref{equ:model_equal}), the output of the equalizer is given by
 	\begin{equation}
	z_n =  S^{H}A_{n}+ C^{H}B_{n} 
	\label{equ:model_matrix_combined}
	\end{equation}
where $S$ denotes the column vector that contains the coefficients of the combined impulse response $s_{l}$ of the channel and equalizer as illustrated in Fig.~\ref{fig:block_diagram}.
Let $w_n$ and $\bar{x}_n$ denote the noise and the residual interference after filtering by the equalizer, respectively. In this paper, our approach differs from work on channel deconvolution because we assume that the residual interference is not null at the equalizer output, $\bar{x}_n\not=0$. So, the output of the equalizer can be written
	\begin{equation}
	z_n = s_{\nu}a_{n-\nu} + \bar{x}_n + w_n
	\label{equ:model_matrix_combined2}
	\end{equation}
where the receiver gain $s_{\nu}$ is defined as $C^{H}He_{\nu}$,  and $e_{\nu}$ denotes the $\nu$th unit row vector (contains only one non zero for a particular value of delay $\nu$).
The Kurtosis of $z_n$ can be expressed by
	\begin{equation}
	K(z)=E\{|z_n|^4\} -2 (E\{|z_n|^2\})^2 - |E\{z_n^2\}|^2
	\label{eq:eq13}
	\end{equation}
where the last term is null when $z_n$ is circular, \textit{e.g.} when $a_n$ is a complex-value so that $E\{a_n^2\}=0$, leading to $E\{z_n^2\}=0$. 
Since symbols $a_n$ are assumed \textit{i.i.d.} and independent of the noise components, the Kurtosis of $z_n$ can be written as a sum of separate Kurtoses
	\begin{equation}
	K(z)=|s_{\nu}|^4 K(a) + K(\bar{x}) + K(w)
	\label{eq:eq14}
	\end{equation}
When the number of taps of the equalizer is sufficiently large, the pdf of the residual interference $\bar{x}_n$ is reasonably well modeled by a Gaussian distribution. Although the ISI distribution for a discrete input can never be Gaussian, it is commonly accepted that the ISI at the output of an MMSE equalizer may be estimate by a Gaussian distribution \cite{Cioffi-1}. This assumption is based on the central limit theorem and is discussed in section \ref{s:stat_gauss}. Thus, since the Kurtosis of a Gaussian random variable is null, we have
	\begin{equation}
	K(z)=|s_{\nu}|^4 K(a)
	\label{eq:eq15}
	\end{equation}
Substituting (\ref{eq:eq15}) in (\ref{eq:eq13}) gives
	\begin{equation}
	|s_{\nu}|^4 K(a) = E\{|z_n|^4\} - 2(E\{|z_n|^2\})^2
	\label{eq:eq16}
	\end{equation}
So, using necessary condition (\ref{eq:eq7}) yields
	\begin{equation}
	|s_{\nu}|^4 K(a) = R_2 E\{|z_n|^2\} - 2(E\{|z_n|^2\})^2
	\label{eq:eq17}
	\end{equation}
The variance of the receiver output $z_n$ can be expressed according to (\ref{equ:model_equal}) as
	\begin{equation}
	E\{|z_n|^2\}=C^{H}R_{yy}C
	\label{equ:Ez2}
	\end{equation}
where $R_{yy}$ is the output autocorrelation matrice of the received data sequence defined by
        \begin{equation}
	R_{yy}=E\{y_{n}y_{n}^{*}\}=\sigma_{b}^{2}I_{M\times M} + \sigma_{a}^{2} HH^{H}
	\label{equ:Ryy}
	\end{equation}
Denote $^{*}$ superscript as the complex conjugate operator. In order to obtain a closed-form for the equalizer $C$, using the gain $s_{\nu}$ we derived the Schwarz inequality which allows us to write 
	\begin{equation}
	|s_{\nu}|^2 =\left| C^{H}He_{\nu}\right|^2 \leq \left(C^{H}R_{yy}C\right)\left(e_{\nu}^{H}H^{H}R_{yy}^{-1}He_{\nu}\right)
       \label{equ:schwartz}
 	\end{equation}
with equality if and only if
      \begin{equation}
	\alpha e_{\nu}^{H}H^{H} = C^{H}R_{yy}
	\label{equ:schwartz_equality}
 	\end{equation}
where $\alpha$ is a complex coefficient. The above inequality follows from \footnote{we consider the unique factorization of the positive-definite matrix $R_{yy}$ in the form $R_{yy}=G^{1/2}G^{H/2}$ and then the decomposition $C^{H}He_{\nu}=C^{H}G^{1/2}G^{-1/2}He_{\nu}$.} (assuming that $R_{yy}$ is invertible).
Let us define
	\begin{equation}
	\Omega_{\nu}= e_{\nu}^{H}H^{H}R_{yy}^{-1}He_{\nu}
	\label{equ:omega}
 	\end{equation}
According to (\ref{equ:Ez2}), expression (\ref{equ:schwartz}) can be rewritten by
	\begin{equation}
	|s_{\nu}|^2 \leq \Omega_{\nu} E\{|z_n|^2\}
	\label{equ:schwartz_equality2}
 	\end{equation}
Considering the sub-Gaussian case ($K(a)<0$), $K(a)=-|K(a)|$. Using (\ref{eq:eq17}) and (\ref{equ:schwartz_equality2}), we show that the necessary condition (\ref{eq:eq7}) to obtain a minimum of the CM cost function involves
	\begin{equation}
	E\{|z_n|^2\} \leq \frac{R_2}{2-|K(a)|\Omega_{\nu}^2}
	\label{eq:eq28}
 	\end{equation}
Substituting now (\ref{eq:eq28}) in (\ref{eq:eq9}), we can see that $\bar{D}^{(2)}$ is lower bounded by
	\begin{equation}
	\bar{D}^{(2)} \geq (R_2)^2 (1-\frac{1}{2-|K(a)|\Omega_{\nu}^2})
	\label{eq:eq29}
	\end{equation}
So we deduce that $\bar{D}^{(2)}$ will be minimum if and only if 
	\begin{equation}
        C^{H}=\alpha e_{\nu}^{H}H^{H} R_{yy}^{-1}	
     	\label{equ:schwartz_equality3}
 	\end{equation}
Since the minimization of $\bar{D}^{(2)}$ with respect to the necessary condition (\ref{eq:eq7}) is equivalent to the minimization of $D^{(2)}$ (section II), the transfer function (\ref{equ:schwartz_equality3}) corresponds to a global minimum. Then requiring the equality of (\ref{eq:eq28}), it results that the output power is bounded by $R_2/2\leq E\{|z_n|^2\}\leq R_2$ which is consistent with \cite{Zeng-6} in the complex case.
To find $\alpha$, we replace $C$ by its value (\ref{equ:schwartz_equality3}) in (\ref{equ:Ez2}) and use (\ref{eq:eq28}). Then, the solution is given by
	\begin{equation}
	|\alpha|^2 = \frac{1}{\Omega_{\nu}} \frac{R_2}{2-|K(a)|\Omega_{\nu}^{2}}
	\label{eq:eq32}
 	\end{equation}
Finally, the finite-length equalizer minimizing the CM criteria is defined up to a phase ambiguity by
	\begin{equation}
	C^{H}=|\alpha| e_{\nu}^{H}H^{H} ( \sigma_{b}^{2}I_{M\times M} + \sigma_{a}^{2} HH^{H}) ^{-1}
        \exp(j\theta), \;\theta \in [0;2\pi]
	\label{equ:sol_cma}
 	\end{equation}
Hence recalling the finite-length MMSE equalizer minimizing the mean square error which is a causal Wiener filter given by
	\begin{equation}
	C_{\textrm{\tiny MMSE}}^{H} = e_{\nu}^{H}H^{H}\sigma_a^2 ( \sigma_{b}^{2}I_{M\times M} + \sigma_{a}^{2} HH^{H}) ^{-1}
	\label{equ:sol_mmse}
 	\end{equation}
And using (\ref{equ:sol_cma}) and (\ref{equ:sol_mmse}), one can easily obtain a relation between the MMSE and CM equalizers
	\begin{equation}
	C_{\textrm{\tiny MMSE}}^{H} = \frac{exp(-j\theta)}{|\alpha| \; \sigma_a^2} \; C^{H}, \;\;\; 	\theta \in [0;2\pi]
	\label{equ_rel_cm_mmse}
 	\end{equation}
Since MMSE and CM equalizers are identical up to a complex factor, it is theoretically possible to reach the MMSE equalizer performance with a blind receiver composed of a CM equalizer followed by a correction gain and a phase rotator, as depicted in Fig.~\ref{fig:comm_model_MMSE}.

\section{ Statistical measure of Gaussian distribution} \label{s:stat_gauss}
The output of the blind and MMSE receiver is shown to be theoretically the same. Consequently, the residual interference at the output of the two receivers will be similar and we can equivalently study the Gaussian nature of the residual interference at the output of the optimum finite length MMSE equalizer. A commonly used approach \cite{Chambers-1983} is to plot the data against a theoretical normal distribution in such a way that if the data set exhibits the properties of a normal distribution the points should lie in a straight line. Deviations reflect miss-matches between the data distribution measured and normal distribution. Specifically if $\bar{x}^{(1)} \leq \bar{x}^{(2)}... \leq \bar{x}^{(N)}$ denote the $N$ ordered observations $\bar{x}_n$, $n=1,...,N$ of the residual interference in (\ref{equ:model_matrix_combined2}) and $q^{(i=1,...,N)}$ the corresponding quantiles, mean the fraction of points below a given value, of the standard normal distribution \cite{Krishnaiah-stat-handbook}, the $N$ points $\left ( q^{(i)},\bar{x}^{(i)}\right)$, $i=1,...,N$ define the normal probability plot.  The distribution of the ISI output is also reported and compared to a theoretical gaussian pdf. We have considered various numbers of equalizer coefficients and results are given by generating N=250000 symbols for two signal to noise ratios (SNR) defined by $SNR=\sigma_a^2 ||h||^{2}/\sigma_b^2$. The channel is time-invariant and given by the following set of real coefficients [0.5679 -0.1136 0.5849 0.1124 0.556]. The distributions are normalized in location and scale.
\begin{figure}[htb]
\begin{tabular}{rcc}
a)&$\hat{C}=0.9988$&$\hat{K}=-0.0037$\\
&\includegraphics[height=4.2cm,width=3.8cm]{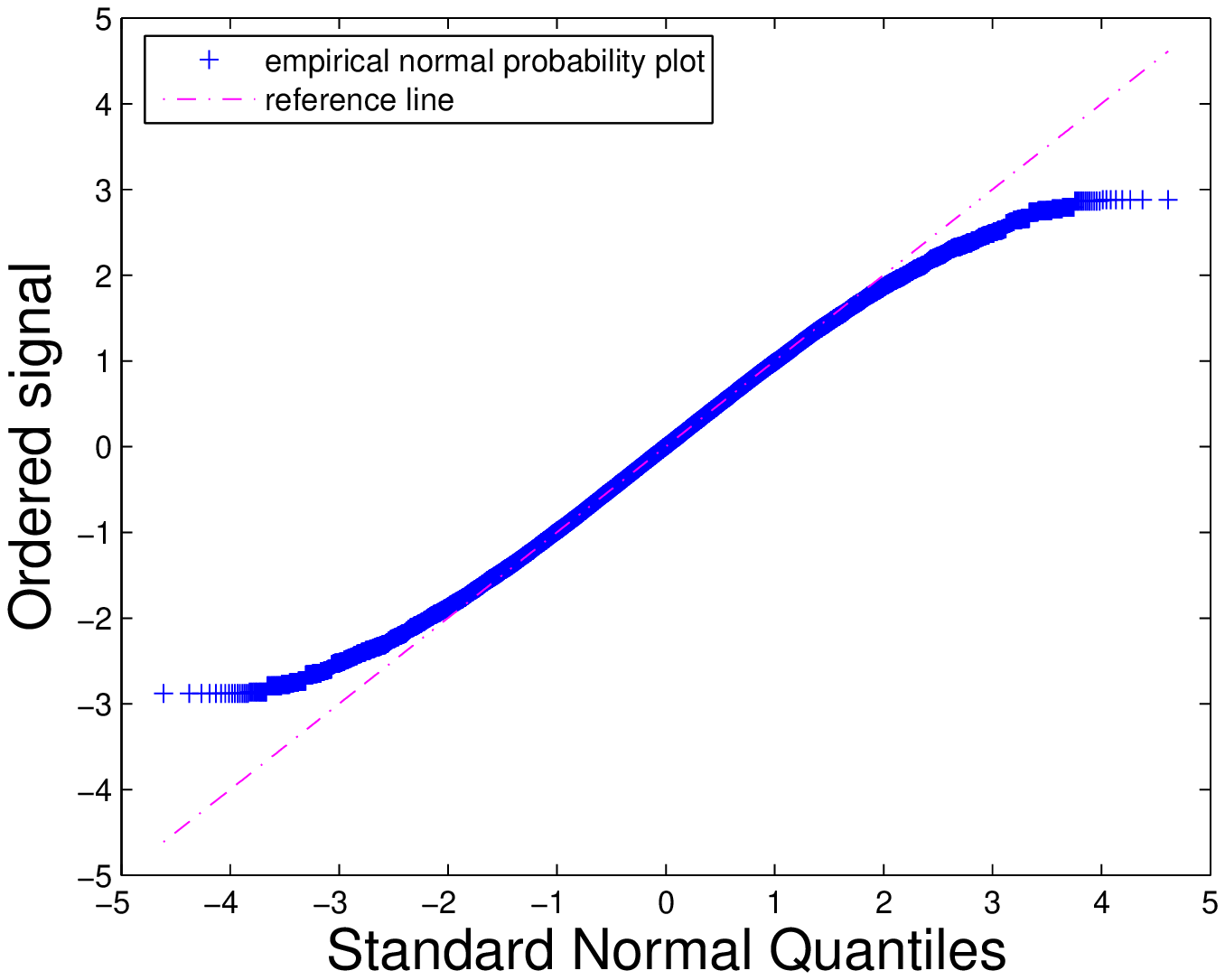}&
\includegraphics[height=4.2cm,width=3.8cm]{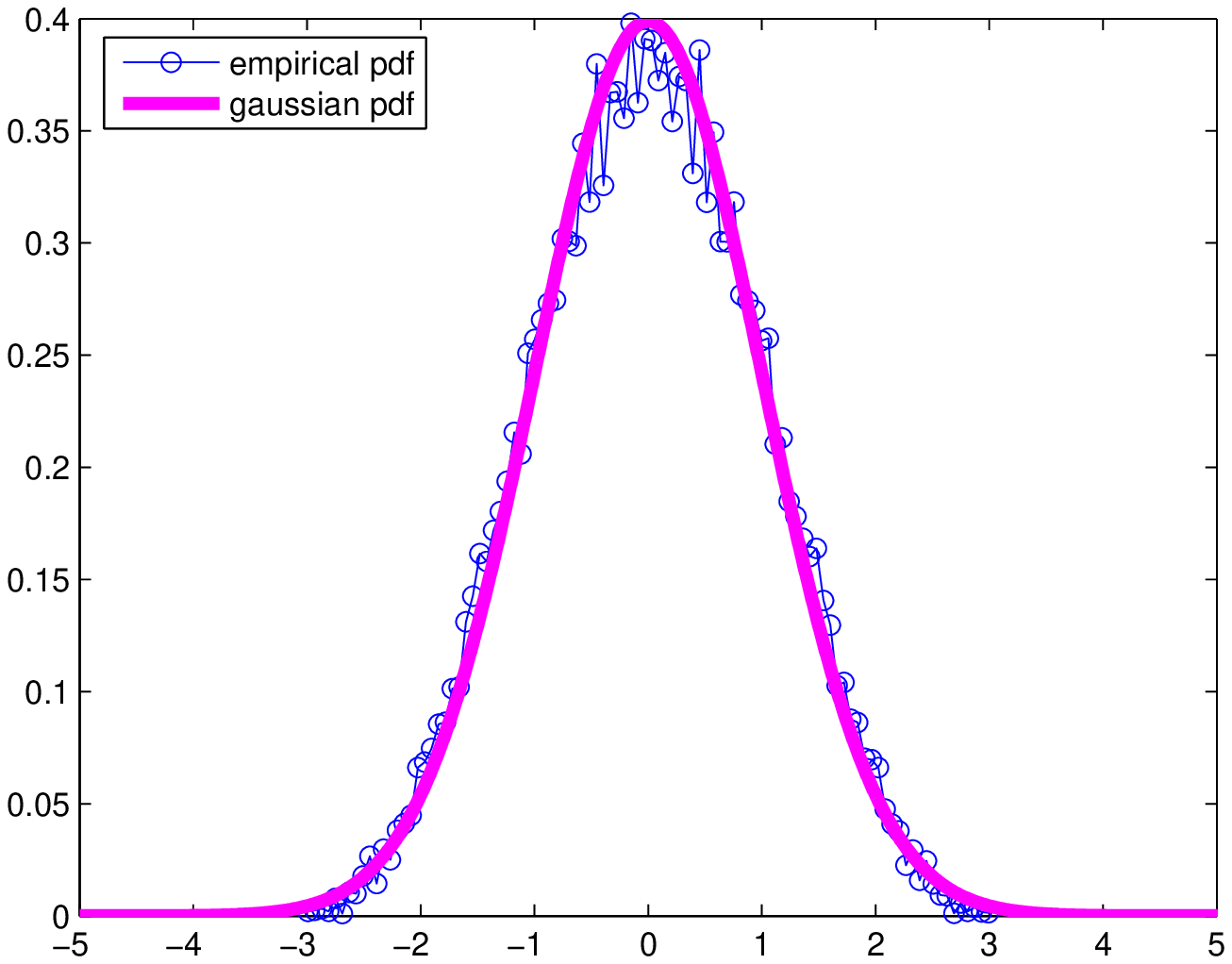}\\
b)&$\hat{C}=0.9997$&$\hat{K}=-0.0003$\\
&\includegraphics[height=4.2cm,width=3.8cm]{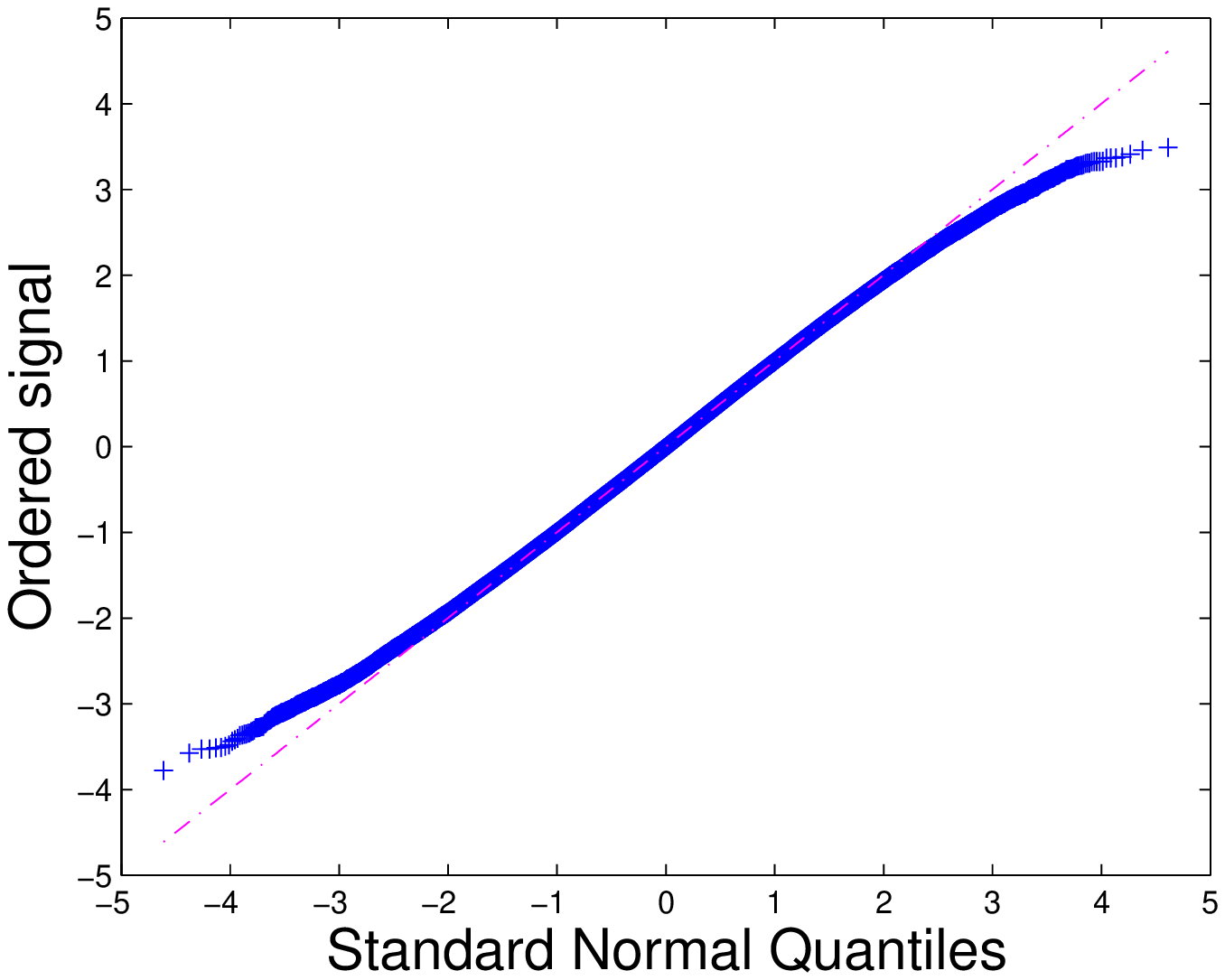}&
\includegraphics[height=4.2cm,width=3.8cm]{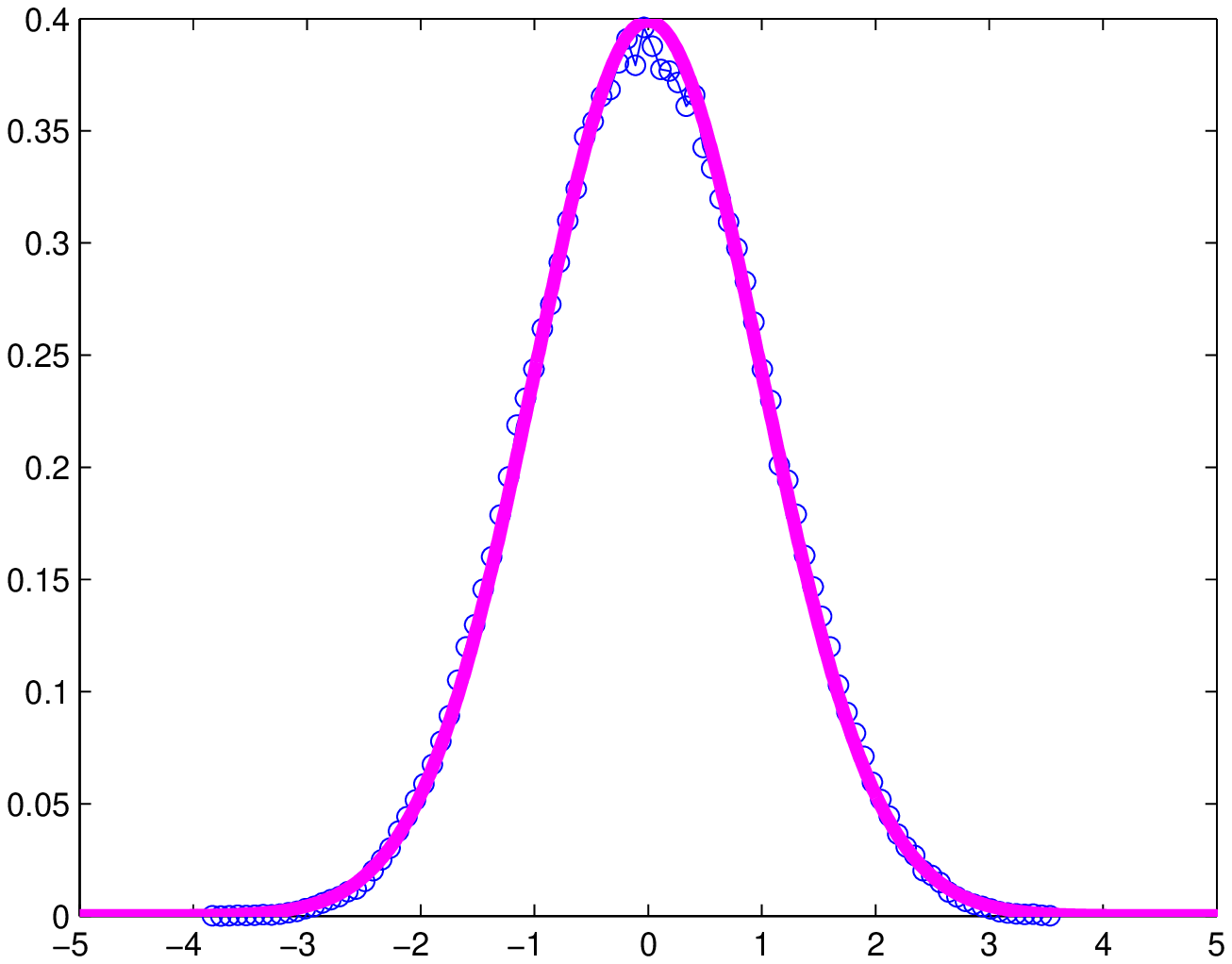}\\
c)&$\hat{C}=0.9999$&$\hat{K}=0$\\
&\includegraphics[height=4.2cm,width=3.8cm]{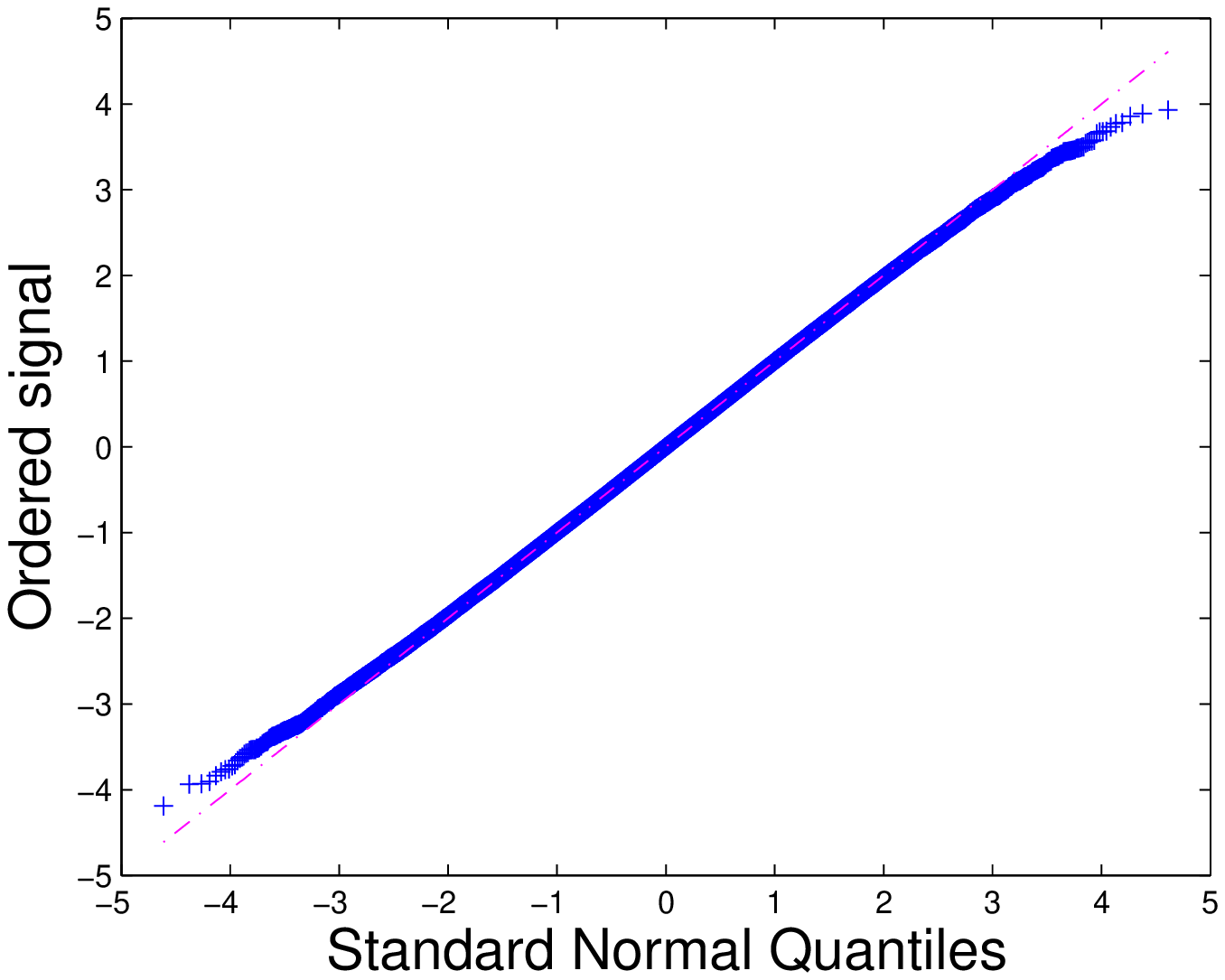}&
\includegraphics[height=4.2cm,width=3.8cm]{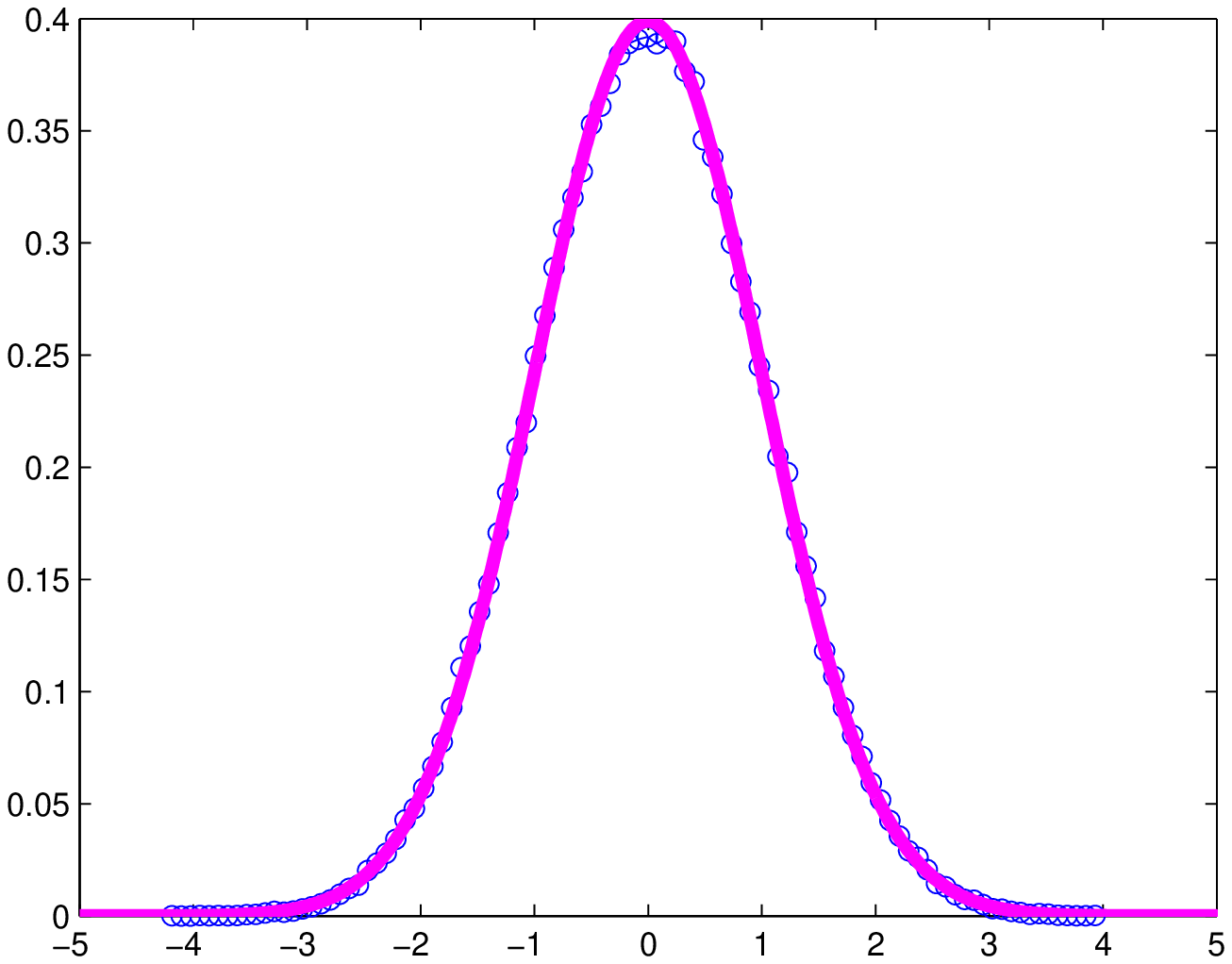}\\
\end{tabular}
\caption{Gaussianity measure. Left: Normal probability plot. The points should fit the reference line to check that the data set is normally distributed. Right: Probability density function. We observe the similarity between the empirical pdf of the data and the theoretical Gaussian pdf. SNR=20dB a) M=11 b) M=21 c) M=41.}\label{fig:stat_gauss_20}\end{figure}
\begin{figure}[rtb]
\begin{tabular}{rcc}
a)&$\hat{C}=0.9989$&$\hat{K}=-0.0034$\\
&\includegraphics[height=4.2cm,width=3.8cm]{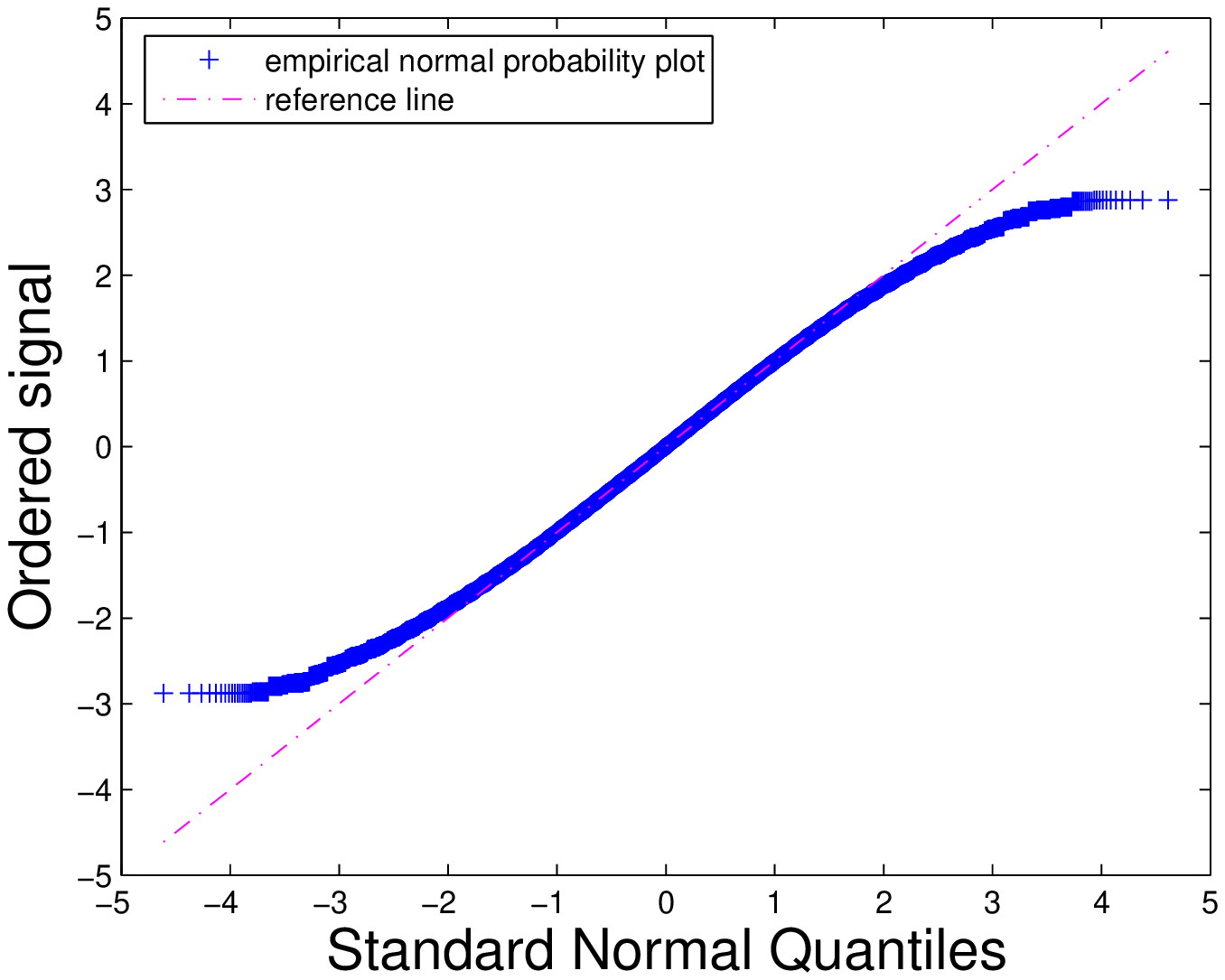}&
\includegraphics[height=4.2cm,width=3.8cm]{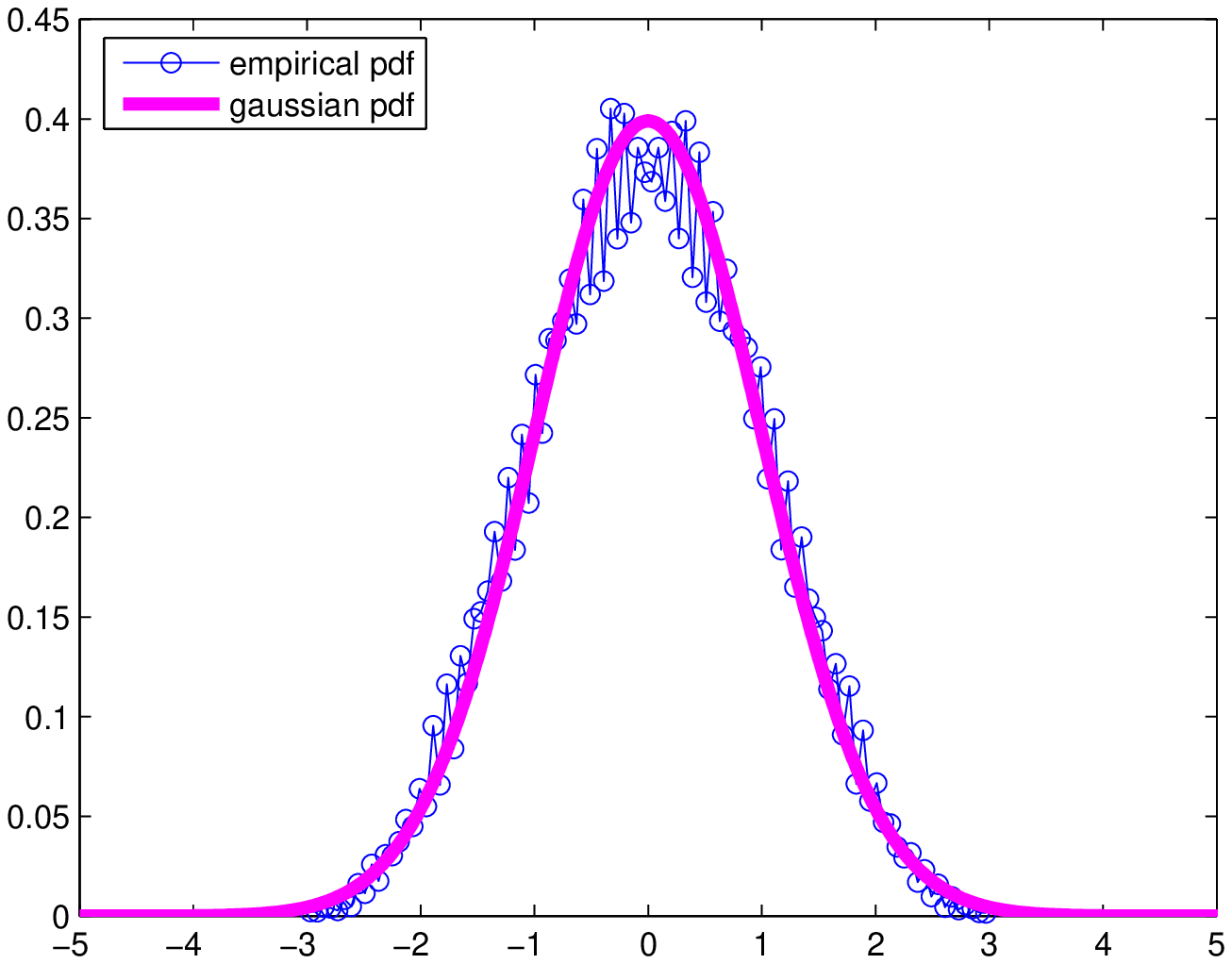}\\
b)&$\hat{C}=0.9997$&$\hat{K}=-0.0003$\\
&\includegraphics[height=4.2cm,width=3.8cm]{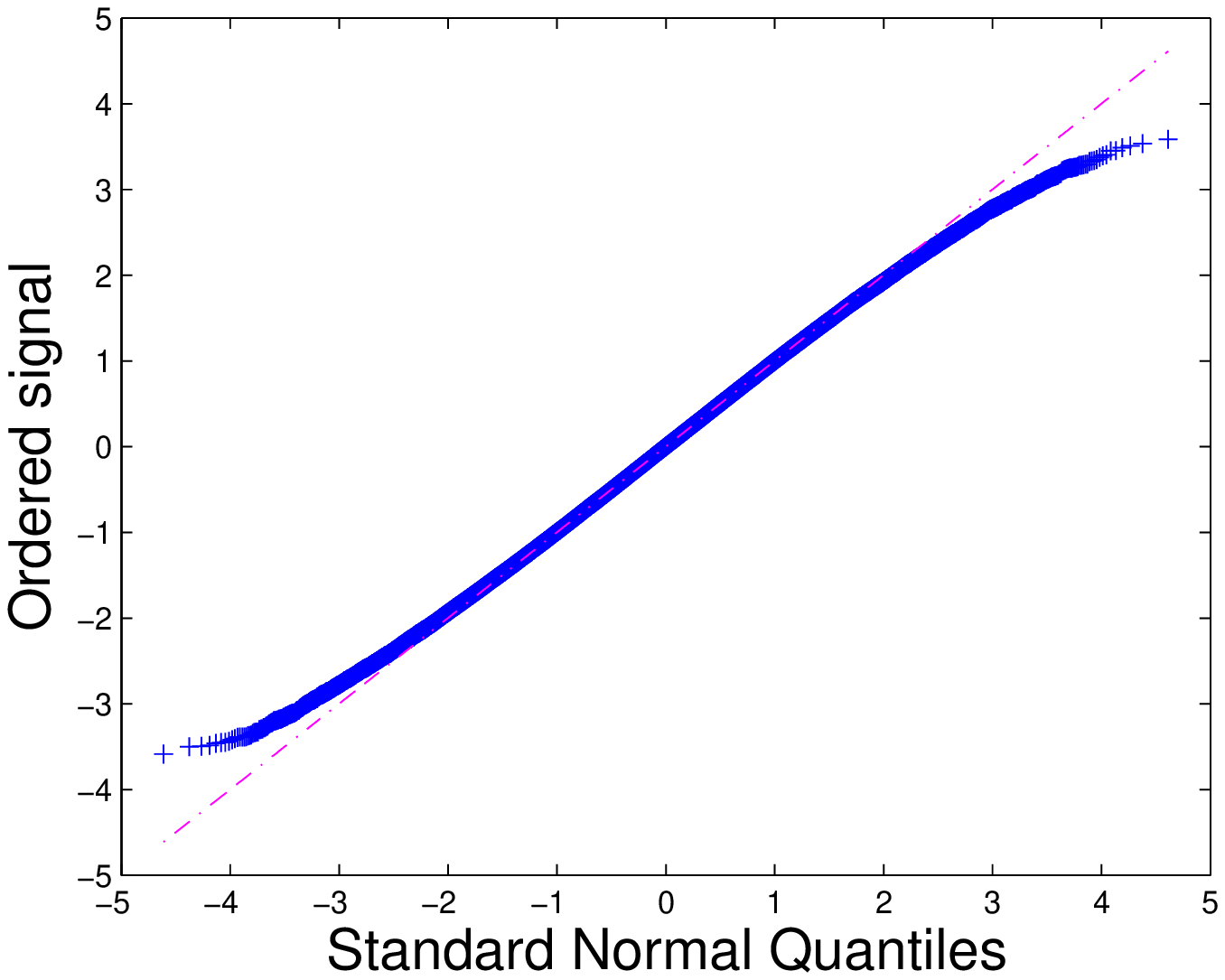}&
\includegraphics[height=4.2cm,width=3.8cm]{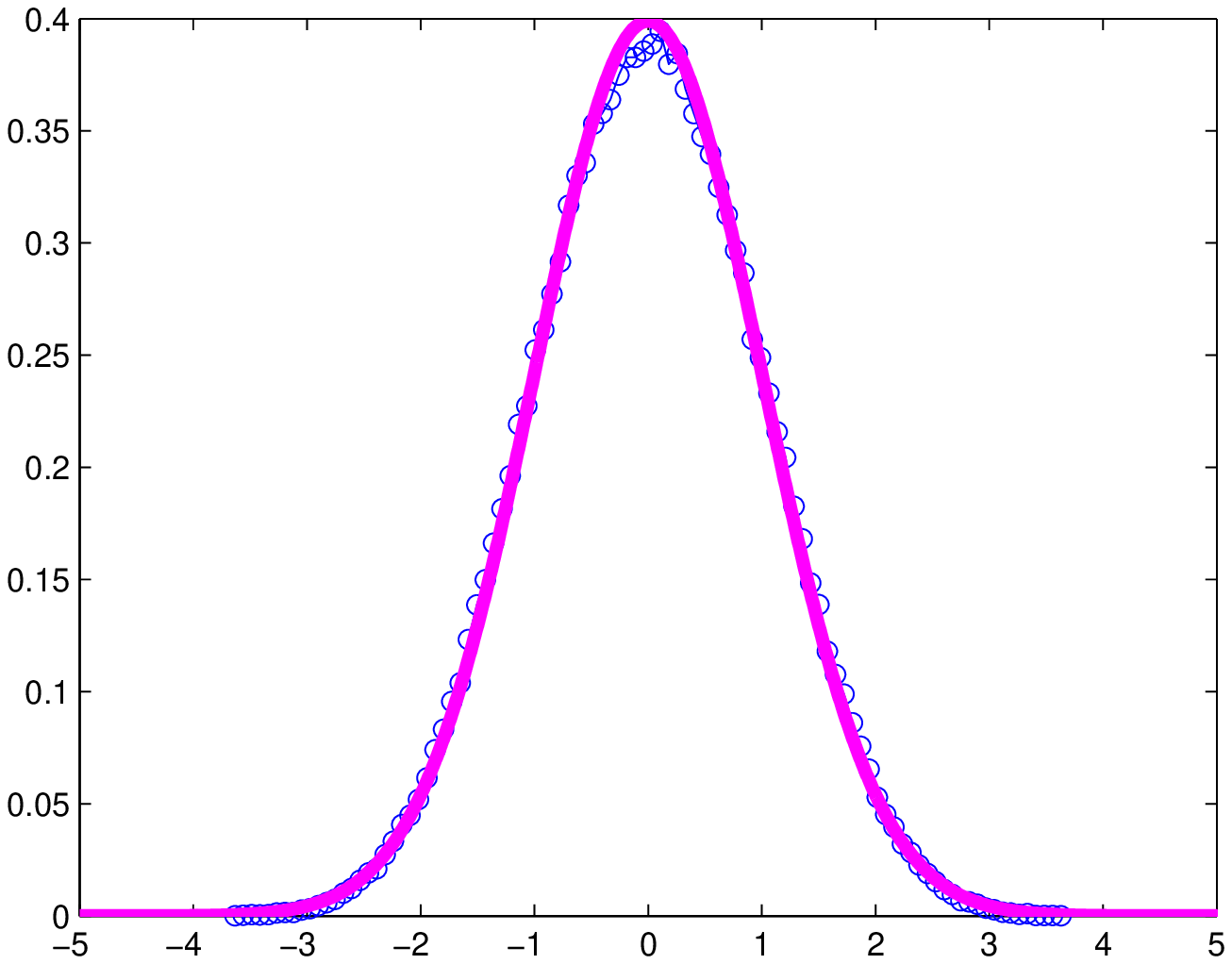}\\
c)&$\hat{C}=0.9997$&$\hat{K}=0$\\
&\includegraphics[height=4.2cm,width=3.8cm]{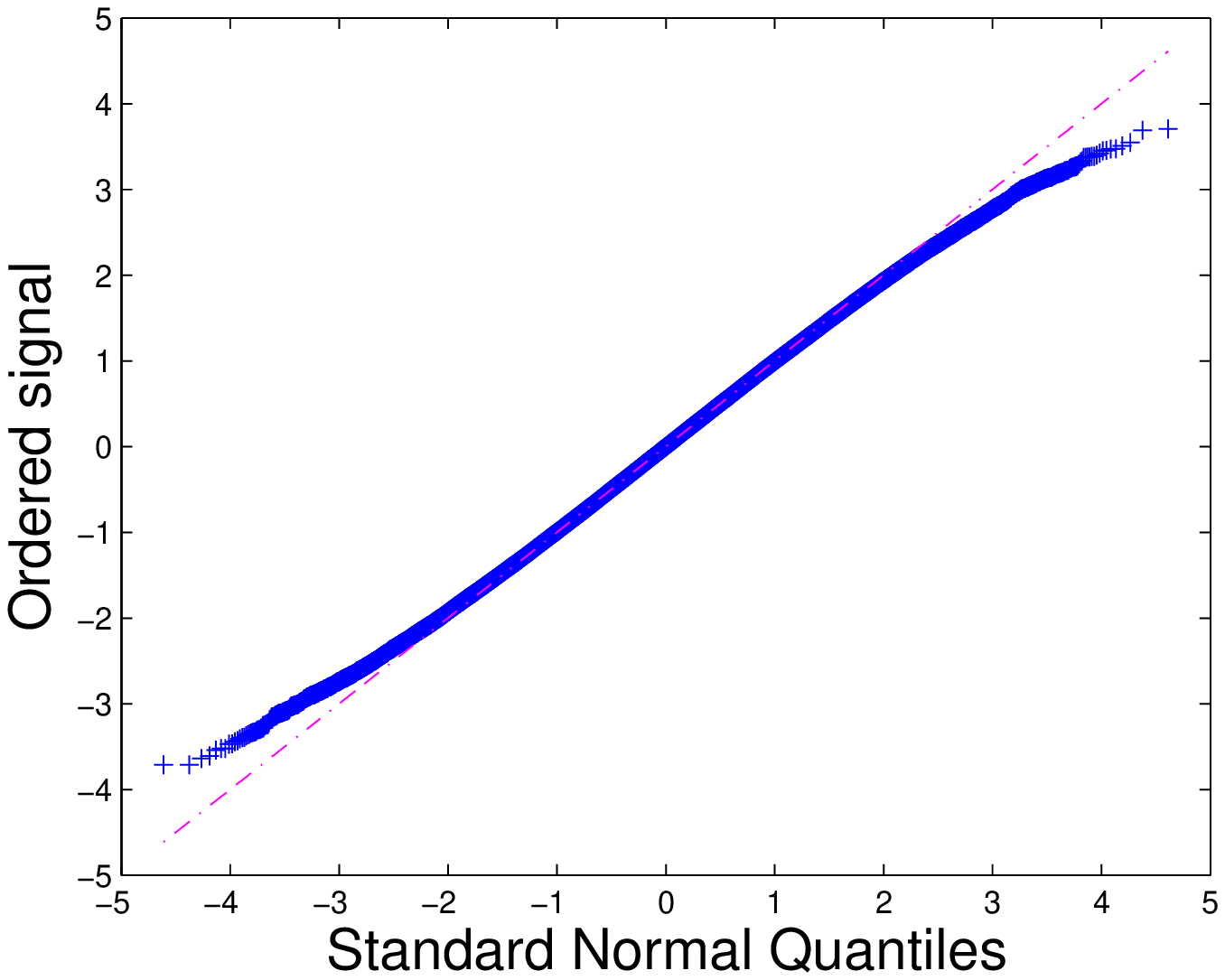}&
\includegraphics[height=4.2cm,width=3.8cm]{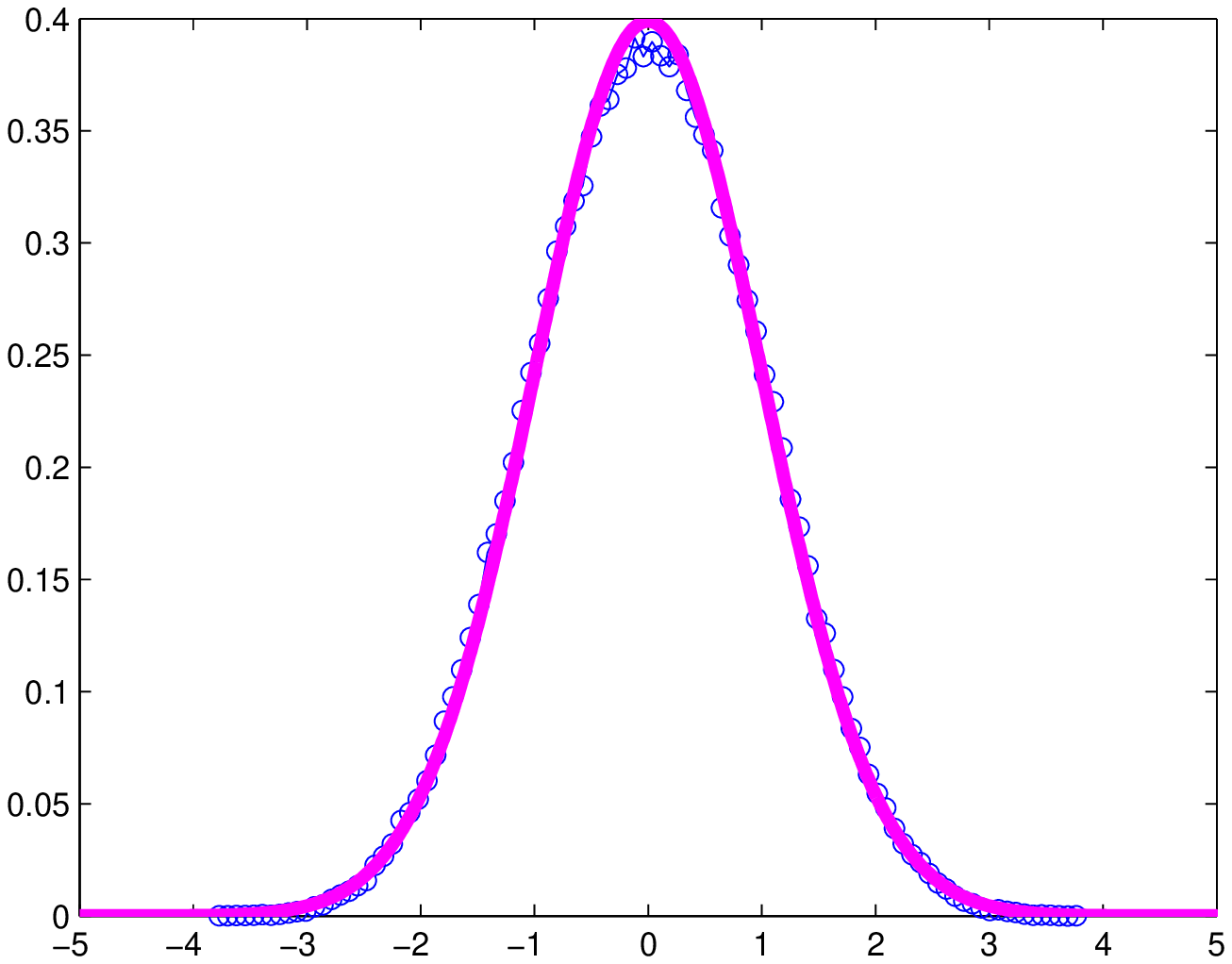}\\
\end{tabular}
\caption{Gaussianity measure. Left: Normal probability plot. The points should fit the reference line to check that the data set is normally distributed. Right: Probability density function. We observe the similarity between the empirical pdf of the data and the theoretical Gaussian pdf. SNR=15dB a) M=11 b) M=21 c) M=41.}\label{fig:stat_gauss_15}\end{figure}
The normal probability plot in Fig.~\ref{fig:stat_gauss_20}-Left shows a strongly linear pattern in the center of the data. The quality of the fit is evaluated through the Pearson correlation coefficient defined by
 \begin{equation}
 \hat{C}(\bar{x},q)=\frac{1}{N}\sum_{i=1}^{N}\bar{x}^{(i)}q^{(i)}
 \label{equ:Pearson_simu}
 \end{equation}
which measures the correlation between the two distributions, a value of +1 indicating a perfect positive linear relationship. We observe that the first and the last few points deviate from the reference line since the number of tap gains $M$ of the equalizer is not large enough. It is noticeable in Fig.~\ref{fig:stat_gauss_20}-Right-a with distortions of the empirical pdf relative to the normal distribution. The same observations are related in Fig.~\ref{fig:stat_gauss_15}-Left / \ref{fig:stat_gauss_15}-Right for an SNR of 15dB. However, in all cases, the tendency towards normality is obvious. A further characterization of the distribution relative to the standard normal one includes the kurtosis measure defined as
\begin{equation}
\hat{K}(\bar{x})=\frac{1}{N}\sum_{n=1}^{N}|\bar{x}_n|^{4}-2\left (\frac{1}{N}\sum_{k=1}^{N}|\bar{x}_n|^{2}\right )^2
\label{equ:kurto_simu}
\end{equation}
As indicated in Fig.~\ref{fig:stat_gauss_20} and Fig.~\ref{fig:stat_gauss_15}, the kurtosis is still near zero which is the key assumption of the task. We can quite conclude that for a reasonable SNR/equalizer length, relation (\ref{eq:eq15}) is achieved.

\section{ Simulation results} \label{s:simu_results}
To confirm our theoretical results, we propose to compare the response of the proposed blind receiver as depicted in Fig.~\ref{fig:comm_model_MMSE}, with that of the Wiener solution. 
If we assume that emitted symbols are known at the receiver, we can use a data-aided least mean square (LMS) algorithm that converges towards the Wiener solution. The tap weight update vector of the equalizer is as follows
	\begin{equation}
	C_{\textrm{\tiny LMS}}(n)= C_{\textrm{\tiny LMS}}(n-1)-\mu(z_n-a_n)Y_n^*
	\label{eq:eq40}
 	\end{equation}
where $\mu$ is a step-size parameter. 
Using a stochastic gradient of the CM cost function (\ref{equ:cost_function_CMeq3}) with respect to the tap weight vector of the blind receiver, each iteration of the algorithm involves updating the following relation
	\begin{equation}
	C(n)= C(n-1)-\mu z_n(|z_n|^2-R_2)Y_n^*
	\label{eq:eq41}
 	\end{equation}
From relation (\ref{equ_rel_cm_mmse}), we need an estimation of $|\alpha|$, denoted $|\hat{\alpha}|$. 
 This value can be obtained blindly from an estimate of the power at the equalizer output. Considering (\ref{eq:eq28}) when the CM cost function is minimum 
	\begin{equation}
	E\{|z_n|^2\}=\frac{R_2}{2-|K(a)|\Omega^2}
	\label{eq:eq42}
 	\end{equation}
from (\ref{eq:eq32}) and (\ref{eq:eq42}), we can write
	\begin{equation}
	|\hat{\alpha}|^2 = \hat{\sigma}_z^2 	\sqrt{\frac{|K(a)|\hat{\sigma}_z^2}{2\hat{\sigma}_z^2 -R_2}}
	\label{eq:eq43}
 	\end{equation}
with $\hat{\sigma}_z^2$ being estimated recursively from the following relation
	\begin{equation}
	\hat{\sigma}_z^2(n) = \lambda\hat{\sigma}_z^2(n-1) + (1-\lambda)|z_n|^2
	\label{eq:eq44}
 	\end{equation}
Finally, the phase rotator can be realized from a second-order carrier-phase tracking loop operating in a decision-directed mode. Simulations consider a 4PSK baseband model where $\sigma_a^2=1$, $K(a)=-1$ and $R_2=1$. The channel is the same set of real coefficients as that of section \ref{s:stat_gauss}. We have considered a high ($M=41$) number of complex tap gains for the equalizers in order to avoid unstable phenomena due to finite-length equalization constraints \cite{Li-5}. The center-tap is initialized to 1 and the step-size $\mu$ is set to 0.001. Results are given after the transmission of 50000 symbols (the algorithm performs one update iteration per symbol). 
\begin{figure}[htb]
\includegraphics[height=4cm,width=8.7cm]{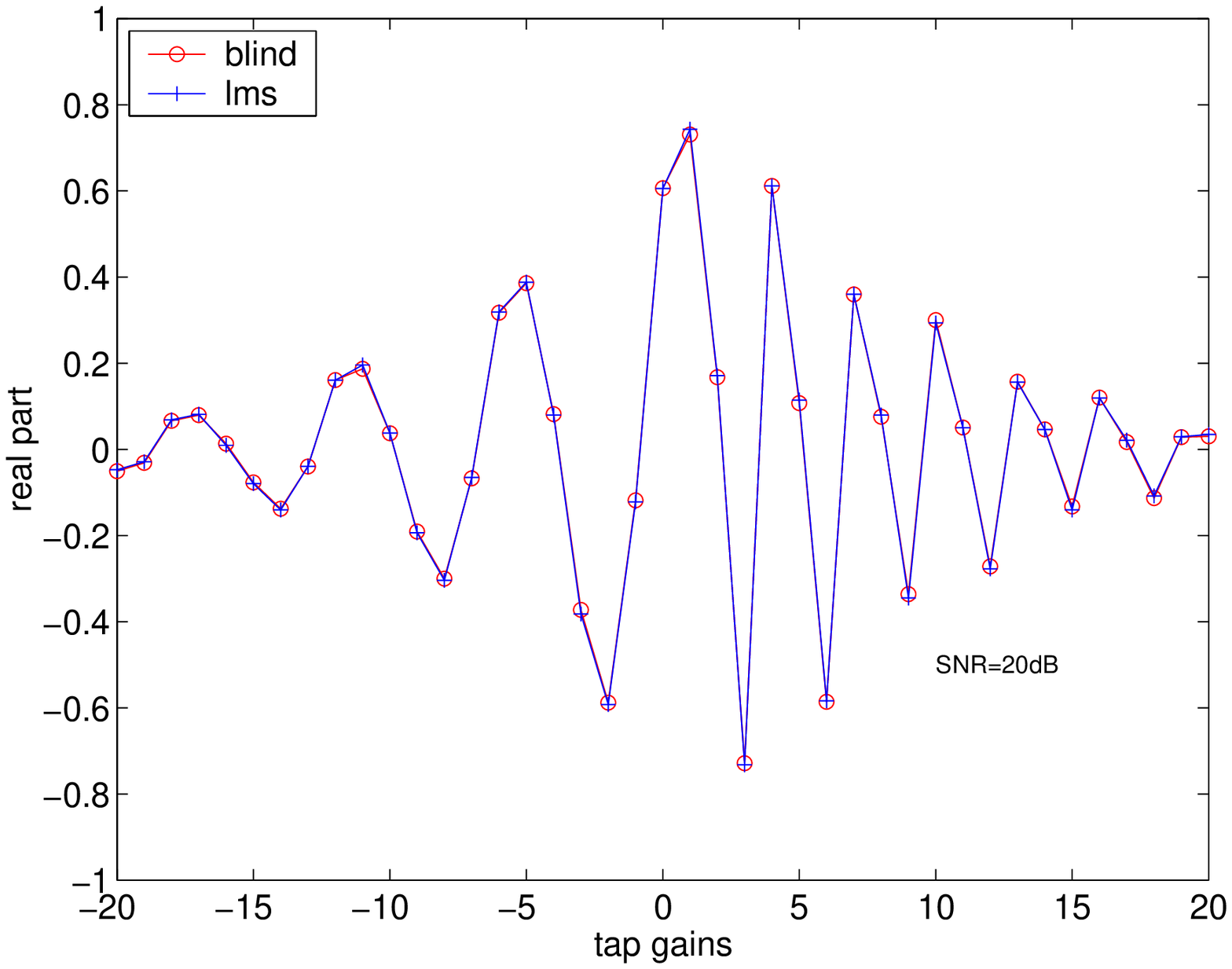}\\
\includegraphics[height=4cm,width=8.7cm]{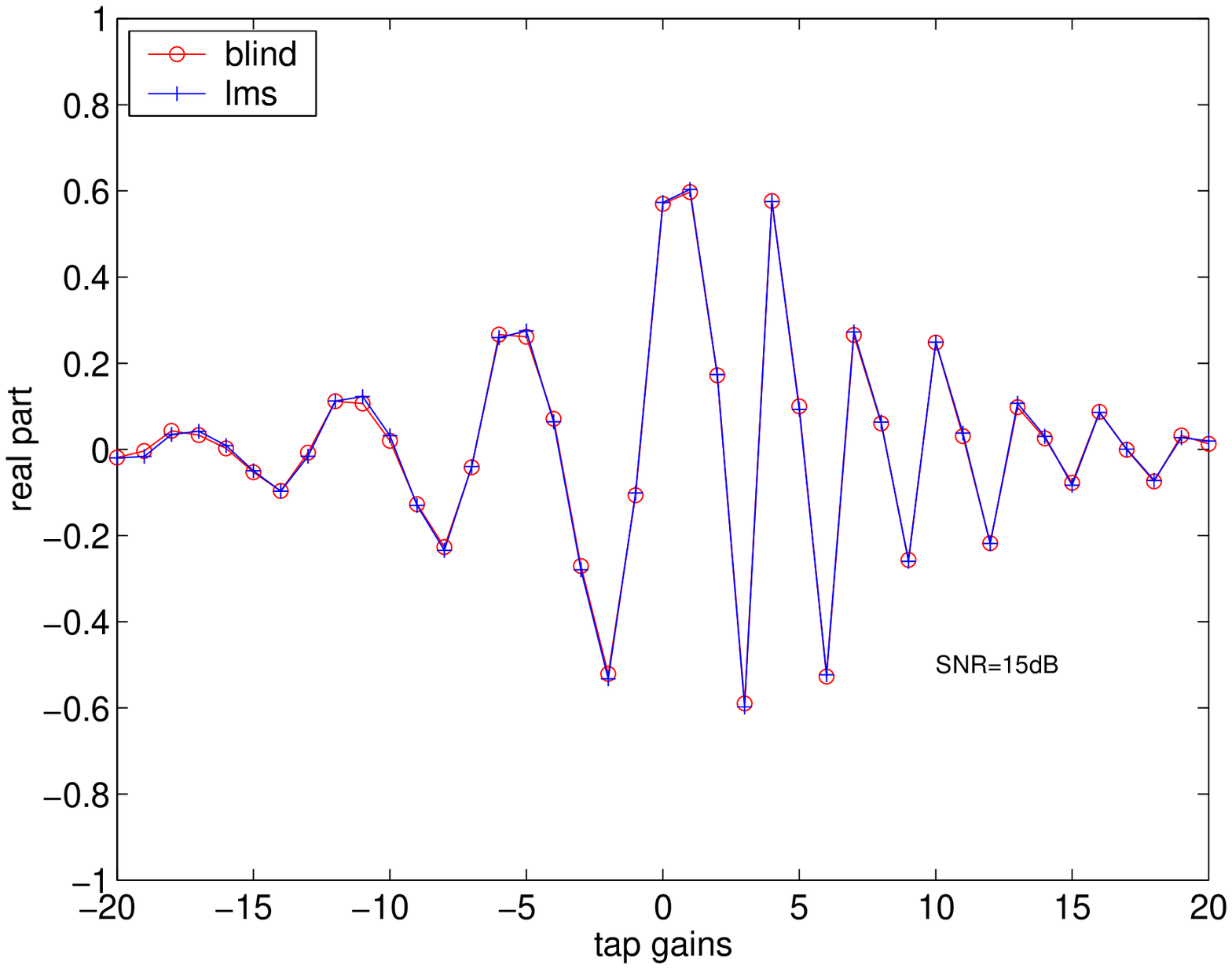}
\caption{Comparison of data-aided (LMS) and blind equalizer coefficients after parameter convergence for SNR=20dB and SNR=15dB. The blind receiver response is very close with that of the MMSE equalizer.}\label{fig:cma_mmse_response}
\end{figure}
As shown in  Fig.~\ref{fig:cma_mmse_response}, the response of the blind receiver including the CM equalizer and the complex gain compensation is very close to that of the LMS equalizer for both SNR. Consequently, the bit error rate performance of the two receivers will be similar. It should be noted that even if asymptotic performance is the same, the data-aided LMS algorithm exhibits faster convergence than the blind algorithm.

\section{Conclusion}
In the present paper a closed-form solution for a finite length CM equalizer under a non-minimum phase channel in the presence of additive white Gaussian noise is presented. The analytical resolution is based on the Gaussian approximation assumption of intersymbol interference at the output of the blind receiver. Statistical measures are used to validate the hypothesis.  In the case of a reasonable length equalizer and finite SNR, the empirical output distribution tends to the normal one and the zero kurtosis assumption is valid. Under this condition, the CM equalizer is shown to be equivalent to the MMSE up to a complex gain factor, which is consistent with the relation between the CM and Wiener receivers proposed by Zeng \textit{et al.} \cite{Zeng-6}. Therefore the CM receiver can theoretically reach the same asymptotic performance as the nonblind receiver. The simulation results herein corroborate the theoretical analysis. Finally we would like to emphasize that the proposed solution clearly deserves further investigation.

\bibliography{ClosedFormCMA_ISIT2005_v3}\bibliographystyle{./ieeetr}

\begin{thebibliography}{1}

\bibitem{Godard-1}
D.~N. {G}odard, ``{S}elf-{R}ecovering {E}qualization and {C}arrier {T}racking
  in {T}wo-{D}imensional {D}ata {C}ommunication {S}ystems.,'' {\em IEEE Trans.
  Commun.}, vol.~28, pp. 1867-1875, Nov. 1980.

\bibitem{Treichler-4}
J.~R. {T}reichler and B.~G. {A}gee, ``{A} {N}ew {A}pproach to {M}ultipath
  {C}orrection of {C}onstant {M}odulus {S}ignals.,'' {\em IEEE Trans. Acoust.,
  Speech, Signal Processing}, vol.~31, NO.2, pp. 459-472, April 1983.

\bibitem{Shalvi-2}
O.~{S}halvi and E.~{W}einstein, ``{N}ew {C}riteria for {B}lind {D}econvolution
  of {N}onminimum {P}hase {S}ystems ({C}hannels).,'' {\em IEEE Trans. Inform.
  Theory}, vol.~36, NO.2, pp. 312-321, March 1990.

\bibitem{Benveniste-3}
M.~G. {A}.~{B}enveniste and G.~{R}uget, ``{R}obust {I}dentification of a
  {N}onminimum {P}hase {S}ystem: {B}lind {A}djustment of a {L}inear {E}qualizer
  in {D}ata {C}ommunications.,'' {\em IEEE Trans. Automat. Contr.}, vol.~25,
  NO.3, pp. 385-399, June 1980.

\bibitem{Zeng-6}
{{H}anks {H}. {Z}eng, {L}ang {T}ong and {R}ichard {J}ohnson Jr.},
  ``{R}elationships {B}etween the {C}onstant {M}odulus and {W}iener
  {R}eceivers.,'' {\em IEEE Trans. Information Theory}, vol.~44, NO.4, pp.
  1523-1538, July 1998.

\bibitem{Cioffi-1}
J.~{C}ioffi, G.~{D}udevoir, M.~{V}edat {Eyuboglu}, and G.~{F}orney, ``{MMSE}
  {D}ecision {F}eedback {E}qualizers and {C}oding - {P}art {I}: {E}qualization
  {R}esults,'' {\em IEEE Trans. Commun.}, vol.~43, NO.10, pp. 2582-2594, Oct.
  1995.

\bibitem{Chambers-1983}
{{J}.{M}.~{C}hambers, {W}.{S}.~{C}leveland, {B}.~{K}leiner and
  {P}.{A}.~{T}ukey}, {\em {{G}raphical {M}ethod for {D}ata {A}nalysis}}.
\newblock 1983.

\bibitem{Krishnaiah-stat-handbook}
{{P}.{R}.~{K}rishnaiah}, {\em {{H}andbook of {S}tatistics 1}}.
\newblock {E}lsevier {S}cience {P}ublishers {B}.{V}., 1980.

\bibitem{Li-5}
{Y}e {L}i and {Z}hi {D}ing, ``{C}onvergence {A}nalysis of {F}inite {L}ength
  {B}lind {A}daptive {E}qualizer.,'' {\em IEEE Trans. Signal Processing},
  vol.~43, NO.9, pp. 2120-2129, Sept. 1995.

\end{thebibliography}

\end{document}